\documentclass[english,prd,english,notitlepage,nofootinbib,preprintnumbers,showkeys]{revtex4}
\usepackage[latin9]{inputenc}
\usepackage{babel}
\usepackage{amsmath}
\usepackage{amssymb}
\usepackage{graphicx}
\usepackage{esint}
\usepackage[unicode=true,pdfusetitle,
 bookmarks=true,bookmarksnumbered=false,bookmarksopen=false,
 breaklinks=false,pdfborder={0 0 1},backref=false,colorlinks=false]
 {hyperref}

\makeatletter

\providecommand{\tabularnewline}{\\}

\@ifundefined{textcolor}{}
{%
 \definecolor{BLACK}{gray}{0}
 \definecolor{WHITE}{gray}{1}
 \definecolor{RED}{rgb}{1,0,0}
 \definecolor{GREEN}{rgb}{0,1,0}
 \definecolor{BLUE}{rgb}{0,0,1}
 \definecolor{CYAN}{cmyk}{1,0,0,0}
 \definecolor{MAGENTA}{cmyk}{0,1,0,0}
 \definecolor{YELLOW}{cmyk}{0,0,1,0}
}

\providecommand{\tabularnewline}{\\}

\@ifundefined{textcolor}{}{%
 \definecolor{BLACK}{gray}{0}
 \definecolor{WHITE}{gray}{1}
 \definecolor{RED}{rgb}{1,0,0}
 \definecolor{GREEN}{rgb}{0,1,0}
 \definecolor{BLUE}{rgb}{0,0,1}
 \definecolor{CYAN}{cmyk}{1,0,0,0}
 \definecolor{MAGENTA}{cmyk}{0,1,0,0}
 \definecolor{YELLOW}{cmyk}{0,0,1,0}
}

\usepackage{babel}

\makeatother

\begin{document}

\title{Deeply virtual meson production on neutrons}

\author{William Brooks, Ivan Schmidt and Marat~Siddikov}

\address{Departamento de Física, Universidad Técnica Federico Santa María,\\
 y Centro Científico - Tecnológico de Valparaíso, Casilla 110-V, Valparaíso,
Chile}

\preprint{USM-TH-356}

\begin{abstract}
In this paper we analyze in detail how the measurements of exclusive
electroproduction of mesons on neutrons would complement the studies
of generalized parton distributions (GPDs) of the proton, providing
independent experimental observables. Some of these processes on neutrons
have very distinctive features, and thus we expect that measurements
on liquid deuterium would allow to clearly distinguish them from similar
processes on protons, giving a very clean probe of the GPD.
In the case of charged meson production, all produced hadrons are
charged, and for this reason we expect that the kinematics of this
process could be easily reconstructed. We estimate the cross-sections
in the kinematics of the Jefferson Laboratory experiments using current
phenomenological GPD models.
\end{abstract}


\maketitle

\section{Introduction}

Understanding the structure of the hadrons nowadays is one of the
major goals of high energy physics, and therefore it occupies a central
place in the program of modern accelerator facilities. Today, this
structure is parametrized in terms of the so-called generalized parton
distributions (GPDs) which can be studied in Bjorken kinematics in
a wide class of processes~\cite{Ji:1998xh,Collins:1998be}. While
the number of processes which can be used for studies of GPDs is rather
large~\cite{Diehl:2003ny,Koempel:2011rc,Duplancic:2018bum,Pire:2017yge,Pedrak:2017cpp,Boussarie:2017umz,Boussarie:2016qop,Lansberg:2015kha,Ivanov:2004vd},
the precision analyses are currently performed with deeply virtual
Compton scattering (DVCS)~\cite{Dupre:2017hfs,Saylor:2018fds,Diehl:2005pc,Belitsky:1997rh,Ji:1996nm,Vanderhaeghen:1999xj,Vanderhaeghen:1998uc,Camacho:2006qlk}
and deeply virtual production of light mesons (DVMP)~\cite{Mueller:1998fv,Ji:1998pc,Radyushkin:1996nd,Radyushkin:1997ki,Radyushkin:2000uy,Collins:1996fb,Brodsky:1994kf,Goeke:2001tz,Diehl:2000xz,Belitsky:2001ns,Belitsky:2005qn,Kubarovsky:2011zz,Anikin:2009bf,Diehl:2007hd,Mankiewicz:1998kg,Mankiewicz:1999tt,Ahmad:2008hp,Goloskokov:2006hr,Goloskokov:2007nt,Kopeliovich:2012dr,Kopeliovich:2013ae,Kopeliovich:2014pea,Siddikov:2016zmt,Schmidt:2015nka},
with most of the existing studies focusing on proton (liquid hydrogen)
targets or production off protons inside nuclei. There are fewer studies
of exclusive processes on neutrons~\cite{Mazouz:2007aa,Mazouz:2017skh,Ho:2018riy,Compton:2017xkt,Bosted:2016leu},
which might be partially due to the technical difficulties with accessing
them experimentally, or perhaps the belief that they probe ``the
same'' GPDs as in case of protons. Given the fact that the amplitude
of hard exclusive process gets contributions from up to a dozen different
GPDs, and that a large number of additional assumptions are involved
in their modeling, we believe that studies on neutron are well justified,
since neutron-induced processes provide independent observables and
can also help to constrain the GPDs of the proton. For the sake of
definiteness, in this paper we will focus on the deeply virtual production
of pions and kaons on neutrons, tacitly implying that such measurements
might be done on liquid deuterium (LD$_{2}$) with minimal uncertainty
from nuclear effects. The feasibility of such measurements (although
at small energies and virtualities too low for consideration in the
Bjorken limit) was recently demonstrated experimentally in~\cite{Mazouz:2017skh,Ho:2018riy,Compton:2017xkt,Bosted:2016leu},
and with the higher energies now available, it is feasible to make
measurements in the Bjorken regime. The deeply virtual meson production
is quite challenging because, as was found experimentally, the cross-section
of this process is dominated by contributions from poorly known transversity
GPDs~\cite{Defurne:2016eiy,Goldstein:2013gra,Goloskokov:2011rd}
convoluted with poorly known twist-three distributions of mesons,
and under the additional assumption that the three-parton distributions
are negligible. While the processes on neutrons also require the use
of model assumptions, we believe that they provide independent observables
which might allow to test the GPDs extracted from analysis of DVMP
on protons. Some of the processes on neutrons have distinctive features,
and for this reason their measurement on liquid deuterium potentially
provides a very clean channel for study of GPDs. As will be explained
in the next section, the neutron GPDs either allow to probe new flavor
combinations, or when contribute to the same combinations as on the
proton, have better sensitivity to the region of negative light-cone
fractions $x$, usually attributed to sea quarks. We also comment
briefly on possible studies of the strange mesons ($K^{+},\,K_{L,S}^{0}$)
production processes, which together with strangeness production on
protons~\cite{Horn:2012zza,Horn:2018ghc,Carmignotto:2018uqj,Goloskokov:2009ia,Goloskokov:2011rd},
might allow for a better understanding of the valence $u$- and $d$-quarks
GPDs: the $SU(3)$ relations~\cite{Frankfurt:1999fp} allow to relate
the nucleon-hyperon transition GPDs to the quark GPDs of the proton,
and the cross-sections of these processes do not get contributions
from gluon GPDs nor from sea quarks (provided sea quarks are flavor
symmetric, as assumed in most parametrizations of the GPDs). 

The paper is organized as follows. In Section~\ref{sec:DVMP_Xsec}
we describe the framework used for the evaluation, and explain in
detail the advantages of neutron-induced processes. The leading twist
contributions are discussed in subsection~\ref{subsec:LeadingTwist},
and in subsection~\ref{subsec:Tw3} we review the corrections due
to transversity GPDs. Finally, in Section~\ref{sec:Results} we present
numerical results using currently available models of GPDs, and draw
conclusions.

\section{Cross-section of the DVMP process}

\label{sec:DVMP_Xsec}As was demonstrated in~\cite{Arens:1996xw,Diehl:2005pc,Diehl:2003ny,Goloskokov:2006hr},
the cross-section for the deeply virtual exclusive meson production,
$\gamma^{*}(q)\,N\left(p_{1}\right)\to\,N'\left(p_{2}\right)\,M$
might be written in the form 
\begin{align}
2\pi\frac{d\sigma}{dt\,d\varphi} & =\left[\epsilon\frac{d\sigma_{L}}{dt\,}+\frac{d\sigma_{T}}{dt}+\sqrt{\epsilon(1+\epsilon)}\cos\varphi\frac{d\sigma_{LT}}{dt}\right.\label{eq:sigma_def-1-1}\\
 & +\left.\epsilon\cos\left(2\varphi\right)\frac{d\sigma_{TT}}{dt}+\sqrt{\epsilon(1+\epsilon)}\sin\varphi\frac{d\sigma_{L'T}}{dt}+\epsilon\sin\left(2\varphi\right)\frac{d\sigma_{T'T}}{dt}\right],\nonumber 
\end{align}
where $\varphi$ is the angle between the lepton scattering and meson
production planes, $t=\left(p_{1}-p_{2}\right)^{2}$ is the invariant
momentum transfer. The cross-sections also depend on the virtuality
$Q^{2}=-q^{2}$ of the intermediate photon, where $q$ is its momentum,
and the Bjorken variable $x_{B}=Q^{2}/2p_{1}\cdot q$. We also used
standard shorthand notations
\[
\epsilon=\frac{1-y-\frac{\gamma^{2}y^{2}}{4}}{1-y+\frac{y^{2}}{2}+\frac{\gamma^{2}y^{2}}{4}}.
\]
for the ratio of transverse and longitudinal photon fluxes, where
\begin{equation}
\gamma=\frac{2\,m_{N}x_{B}}{Q},\quad y=\frac{Q^{2}}{s_{ep}\,x_{B}}=\frac{Q^{2}}{2m_{N}E_{e}\,x_{B}}.\label{eq:elasticity}
\end{equation}

The cross-section of the subprocess $\gamma^{*}n\to M\,n'$ is related
to the partial amplitudes as~\cite{Arens:1996xw,Goloskokov:2009ia}
\begin{align}
\frac{d\sigma_{L}}{dt\,} & =\Gamma\sigma_{00},\label{eq:sigma_L-1}\\
\frac{d\sigma_{T}}{dt} & =\Gamma\,\left(\frac{\sigma_{++}+\sigma_{--}}{2}+r_{L}\sqrt{1-\epsilon^{2}}\frac{\sigma_{++}-\sigma_{--}}{2}\right),\label{eq:sigma_T-1}\\
\frac{d\sigma_{LT}}{dt} & =-\Gamma\,\left({\rm Re}\left(\sigma_{0+}-\sigma_{0-}\right)+r_{L}\sqrt{\frac{1-\epsilon}{1+\epsilon}}{\rm Re}\left(\sigma_{0+}+\sigma_{0-}\right)\right),\label{eq:sigma_LT-1}\\
\frac{d\sigma_{TT}}{dt} & =-\Gamma\,{\rm Re}\left(\sigma_{+-}\right),\label{eq:sigma_TT-1}\\
\frac{d\sigma_{L'T}}{dt} & =\Gamma\,\left({\rm Im}\left(\sigma_{+0}+\sigma_{-0}\right)+r_{L}\sqrt{\frac{1-\epsilon}{1+\epsilon}}{\rm Im}\left(\sigma_{-0}-\sigma_{+0}\right)\right),\label{eq:LPrimeT}\\
\frac{d\sigma_{T'T}}{dt} & =\Gamma\,{\rm Im}\left(\sigma_{+-}\right),\label{eq:TPrimeT}\\
\Gamma & =\frac{1}{32\pi\left(W^{2}-m^{2}\right)\Lambda\left(W^{2},-Q^{2},\,m^{2}\right)}
\end{align}
where $r_{L}$ is the polarization of the lepton beam, $\Lambda$
stands for the Mandelstam function $\Lambda(x,\,y,\,z)=\sqrt{x^{2}+y^{2}+z^{2}-2xy-2xz-2yz}$,
and the subindices $\alpha,\beta$ of the matrix $\sigma$ refer to
the polarization states of the intermediate photon in the amplitude
and its conjugate. The matrix $\sigma_{\alpha\beta}$ is related to
the helicity amplitudes $\mathcal{A}_{\nu'0,\nu\beta},$ by 
\begin{equation}
\sigma_{\alpha\beta}=\sum_{\nu\nu'}\mathcal{A}_{\nu'0,\nu\alpha}^{*}\mathcal{A}_{\nu'0,\nu\beta},
\end{equation}
where $\nu,\,\nu'$ are the polarization subindices of the initial
and final hadrons. The amplitudes $\mathcal{A}$ carry all the information
about the structure of the hadron. It is expected that in the formal
Bjorken limit ($Q^{2}\to\infty,\,x_{{\rm B}}={\rm const}$) the dominant
contribution comes from the leading twist term, $\sigma_{00}$, whereas
all the other contributions should be suppressed at least as $\sim\mathcal{O}\left(m_{N}/Q\right)$.
However, as was found in~\cite{Goloskokov:2011rd}, in the JLab kinematics
we are far from this regime, and the contributions of other harmonics
in certain channels might yield contributions on par with the leading
twist result. In the following subsections~(\ref{subsec:LeadingTwist},~\ref{subsec:Tw3})
we discuss in detail the contributions for the leading and subleading
twists and the information on GPDs which they carry. In what follows
we will consider only the case of unpolarized beams~($r_{L}=0$)
and targets, since the corresponding asymmetries might be observed
only if the target is polarized, and in the case of neutrons inside
a nucleus this presents a difficult technical problem~\cite{Zheng:2004ce,Huang:2011bc,Flay:2016wie,Airapetian:2018rlq}
and requires modeling of nuclear interactions~\cite{RondonAramayo:1999da}.
In this limit, the cross-sections $d\sigma_{L'T}$ (\ref{eq:LPrimeT})
and $d\sigma_{T'T}$ ~(\ref{eq:TPrimeT}) vanish. 

\subsection{Leading twist contribution}

\label{subsec:LeadingTwist}
\begin{figure}[htp]
\includegraphics[width=9cm]{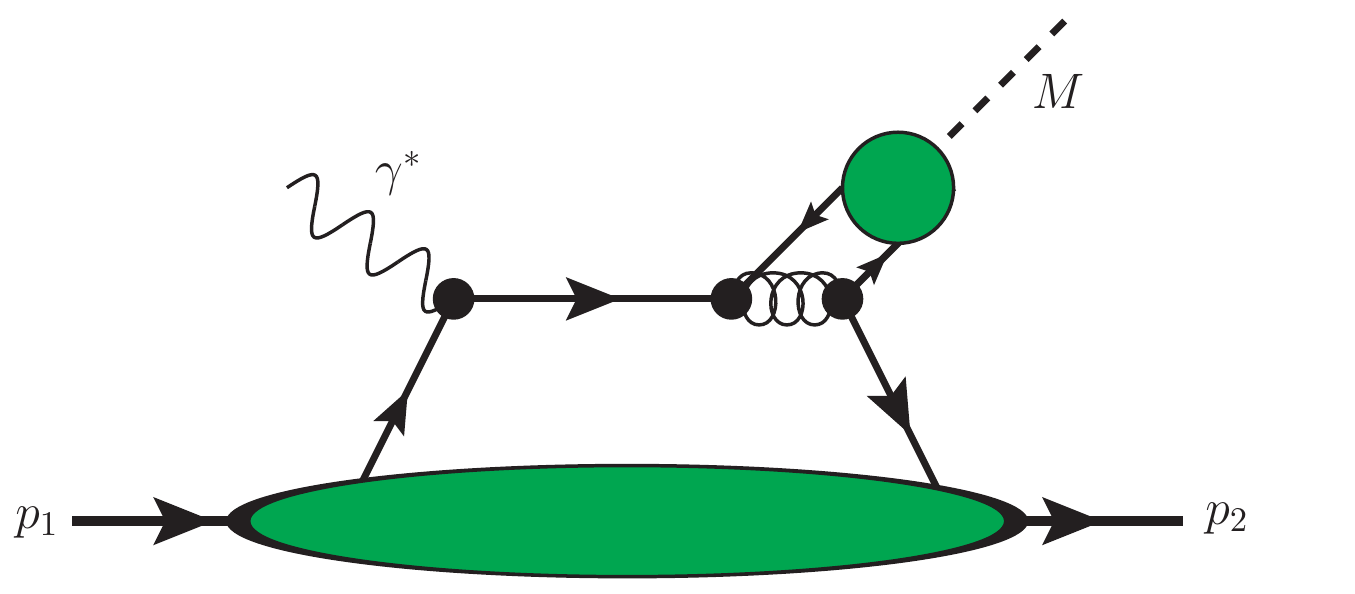}\includegraphics[width=9cm]{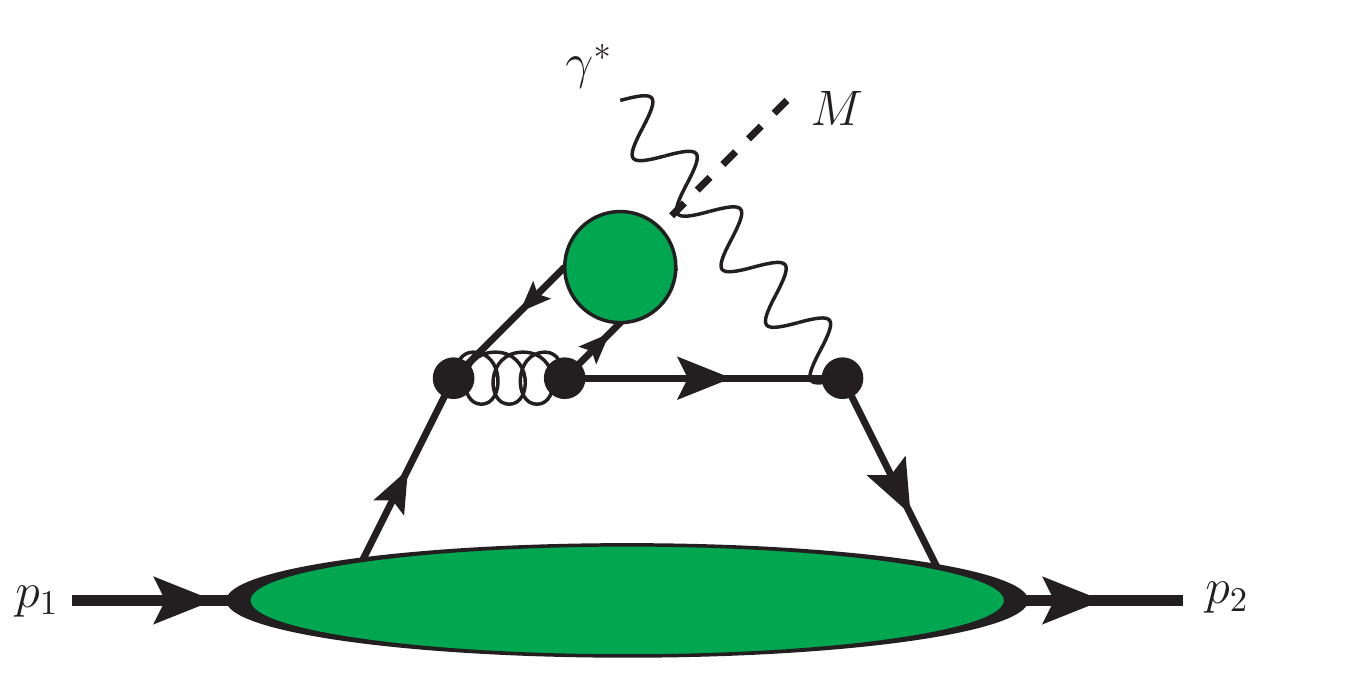}\protect\caption{\label{fig:DVMPLO}Leading-order contributions to the DVMP hard coefficient
functions, $s$-channel (left) and $u$-channel (right). The horizontal
green blob stands for the generalized parton distributions of the
parton, and the upper (small) green blob is the wave function of the
produced meson. In the next-to-leading order, we should add additional
gluon in all possible ways to the upper part of the diagram. The two-parton
twist-three contributions have the same structure.}
\end{figure}

The amplitude of the physical process
in the formal Bjorken limit ($Q^{2}\to\infty$) factorizes into convolution
of the hard and the soft parts, as shown in Figure~\ref{fig:DVMPLO}
\begin{equation}
\mathcal{A}_{\nu',\nu}\left(\xi,\,t\right)=\int_{-1}^{+1}dx\sum_{q}\,\mathcal{H}_{\nu'\lambda',\nu\lambda}^{q}\left(x,\,\xi,\,t,\,\mu_{F}\right)\mathcal{C}_{\lambda\lambda'}^{q}\left(x,\,\xi,\,\mu_{F}\right),\,\label{eq:amplitude}
\end{equation}
where the sum runs over all parton flavors; $\lambda,\,\lambda'$
are helicities of partons, $\nu,\,\nu'$ are the helicities of the
initial and final hadron, the skewness $\xi$ is related to light-cone
momenta of the proton $p_{1,2}$ before and after interaction as $\xi=\left(p_{1}^{+}-p_{2}^{+}\right)/\left(p_{1}^{+}+p_{2}^{+}\right)\approx x_{B}/(2-x_{B})$,
$\mu_{F}$ is the factorization scale, and all the other variables
were defined earlier (see e.g.~\cite{Goeke:2001tz,Diehl:2003ny}
for details of the kinematics). The soft matrix elements $\mathcal{H}^{q}$
in~(\ref{eq:amplitude}) is diagonal in quark helicities ($\lambda,\,\lambda'$)
at leading twist, and can be parametrized in terms of four quark GPDs
$H,\,E,\,\tilde{H},\,\tilde{E}$ as 
\begin{align}
\mathcal{H}_{\nu'\lambda',\nu\lambda}^{q} & =\frac{2\delta_{\lambda\lambda'}}{\sqrt{1-\xi^{2}}}\left(-\left(\begin{array}{cc}
\left(1-\xi^{2}\right)H^{q}-\xi^{2}E^{q} & \frac{\left(\Delta_{1}+i\Delta_{2}\right)E^{q}}{2m}\\
-\frac{\left(\Delta_{1}-i\Delta_{2}\right)E^{q}}{2m} & \left(1-\xi^{2}\right)H^{q}-\xi^{2}E^{q}
\end{array}\right)_{\nu'\nu}\right.\label{eq:HAmp-1}\\
 & +\left.{\rm sgn}(\lambda)\left(\begin{array}{cc}
-\left(1-\xi^{2}\right)\tilde{H}^{q}+\xi^{2}\tilde{E}^{q} & \frac{\left(\Delta_{1}+i\Delta_{2}\right)\xi\tilde{E}^{q}}{2m}\\
\frac{\left(\Delta_{1}-i\Delta_{2}\right)\xi\tilde{E}^{q}}{2m} & \left(1-\xi^{2}\right)\tilde{H}^{q}-\xi^{2}\tilde{E}^{q}
\end{array}\right)_{\nu'\nu}\right),\nonumber 
\end{align}

For the processes in which the baryon state changes, e.g. $en\to e\pi^{-}p$,
the transition GPDs are linearly related via the $SU(3)$ relations~\cite{Frankfurt:1999fp}
to ordinary GPDs. The so-called coefficient functions $\mathcal{C}^{q}$
in~(\ref{eq:amplitude}) are the parton-level amplitudes and are
evaluable in perturbative QCD. They might be represented as a sum
of the $s$- and $u$-channel contributions,
\begin{equation}
\mathcal{C}^{q}\left(x,\,\xi\right)=\mathcal{C}_{s{\rm -channel}}^{q}\left(x,\,\xi\right)+\mathcal{C}_{u{\rm -channel}}^{q}\left(x,\,\xi\right).
\end{equation}
as shown in Figure~\ref{fig:DVMPLO}. In the Bjorken limit, these
functions have an extremely simple form,

\begin{equation}
\left.\begin{array}{c}
\mathcal{C}_{s{\rm -channel}}^{q({\rm LO})}\\
\mathcal{C}_{u{\rm -channel}}^{q({\rm LO})}
\end{array}\right\} =\eta_{q}^{(\mp)}c_{\mp}^{(q)}\left(x,\,\xi\right)+\mathcal{O}\left(\frac{m^{2}}{Q^{2}}\right)+\mathcal{O}\left(\alpha_{s}^{2}\left(\mu_{R}^{2}\right)\right),
\end{equation}
where $\eta_{q}^{(\mp)}$ are some process-dependent flavor factors,
and we introduced a shorthand notation

\begin{eqnarray}
c_{\pm}^{(q)}\left(x,\xi\right) & = & \left(\int dz\frac{\phi_{2,M}\left(z\right)}{z}\right)\frac{8\pi i}{9}\frac{\alpha_{s}\left(\mu_{R}^{2}\right)f_{M}}{Q}\frac{1}{x\pm\xi\mp i0}\left(1+\frac{\alpha_{s}\left(\mu_{r}^{2}\right)}{2\pi}T^{(1)}\left(\frac{\xi\pm x}{2\xi},\,z\right)\right),\label{eq:c2-1}
\end{eqnarray}
where $\phi_{2,\,M}(z)$ is the twist-2 distribution amplitude (DA)
of the produced meson~\cite{Kopeliovich:2011rv}; the NLO correction~$T^{(1)}$
was evaluated in~\cite{Belitsky:2001nq,Ivanov:2004zv,Diehl:2007hd,Braaten:1987yy,Melic:1998qr}
and for the sake of completeness is given explicitly in Appendix~\ref{sec:NLOCoef}.
While the coefficient function is known up to NLO ~\cite{Belitsky:2001nq,Ivanov:2004zv,Diehl:2007hd,Braaten:1987yy,Melic:1998qr},
currently there is no detailed NLO GPDs (especially taking into account
strange quarks) available from the literature. For this reason we
will stick to the LO expressions. From the structure of~(\ref{eq:c2-1})
we may see that the $s$-channel amplitude probes the GPDs near the
point $x\approx\xi\approx x_{B}/(2-x_{B})$, a region dominated by
the valence quarks. At the same time, after trivial algebraic rewrite,
\begin{equation}
\int_{-1}^{+1}dx\,\mathcal{H}^{(q)}\left(x,\,\xi\right)c_{+}^{(q)}\left(x,\,\xi\right)=-\int_{-1}^{+1}dx\,\mathcal{H}^{(q)}\left(-x,\,\xi\right)c_{-}^{(q)}\left(x,\,\xi\right)\label{eq:MConv}
\end{equation}
we observe that the $u$-channel amplitude might be interpreted as
a probe of the GPDs at \emph{negative} $x\approx-\xi$, the region
dominated by sea quarks. The values of the flavor factors are summarized
in Table~\ref{tab:flavourFactors} and determine the sensitivity
of each process to valence and sea quark GPDs. 

\begin{table}
\renewcommand\arraystretch{2}

\begin{tabular}{|c|c|c|c|c|c|c|}
\cline{1-3} \cline{5-7} 
\textbf{Process}  & \textbf{$\eta_{+}^{q}$} & \textbf{$\eta_{-}^{q}$} &  & \textbf{Process}  & \textbf{$\eta_{+}^{q}$} & \textbf{$\eta_{-}^{q}$}\tabularnewline
\cline{1-3} \cline{5-7} 
$e\,n\to e\,\pi^{-}p$  & $e_{u}\left(\delta_{qu}-\delta_{qd}\right)$ & $e_{d}\left(\delta_{qu}-\delta_{qd}\right)$ &  & $e\,p\to e\,\pi^{+}n$  & $e_{d}\left(\delta_{qu}-\delta_{qd}\right)$ & $e_{u}\left(\delta_{qu}-\delta_{qd}\right)$\tabularnewline
\cline{1-3} \cline{5-7} 
$e\,n\to e\,\pi^{0}n$  & $\frac{e_{u}\delta_{qd}-e_{d}\delta_{qu}}{\sqrt{2}}$ & $\frac{e_{u}\delta_{qd}-e_{d}\delta_{qu}}{\sqrt{2}}$ &  & $e\,p\to e\,\pi^{0}p$  & $\frac{e_{u}\delta_{qu}-e_{d}\delta_{qd}}{\sqrt{2}}$ & $\frac{e_{u}\delta_{qu}-e_{d}\delta_{qd}}{\sqrt{2}}$\tabularnewline
\cline{1-3} \cline{5-7} 
\multicolumn{1}{c}{} & \multicolumn{2}{c}{} & \multicolumn{1}{c}{} & \multicolumn{1}{c}{} & \multicolumn{2}{c}{}\tabularnewline
\cline{1-3} \cline{5-7} 
$e\,n\to eK^{0}\Lambda$  & $-e_{s}\frac{2\delta_{qd}-\delta_{qu}-\delta_{qs}}{\sqrt{6}}$ & $-e_{d}\frac{2\delta_{qd}-\delta_{qu}-\delta_{qs}}{\sqrt{6}}$ &  & $e\,p\to e\,K^{+}\Lambda$  & $-e_{s}\frac{2\delta_{qu}-\delta_{qd}-\delta_{qs}}{\sqrt{6}}$ & $-e_{u}\frac{2\delta_{qu}-\delta_{qd}-\delta_{qs}}{\sqrt{6}}$\tabularnewline
\cline{1-3} \cline{5-7} 
$e\,n\to eK^{0}\Sigma_{0}$  & $-e_{s}\frac{\delta_{qu}-\delta_{qs}}{\sqrt{2}}$ & $-e_{d}\frac{\delta_{qu}-\delta_{qs}}{\sqrt{2}}$ &  & $e\,p\to e\,K^{+}\Sigma_{0}$  & $e_{s}\frac{\delta_{qd}-\delta_{qs}}{\sqrt{2}}$ & $e_{u}\frac{\delta_{qd}-\delta_{qs}}{\sqrt{2}}$\tabularnewline
\cline{1-3} \cline{5-7} 
$e\,n\to e\,K^{+}\Sigma^{-}$  & $-e_{s}\left(\delta_{qu}-\delta_{qs}\right)$ & $-e_{u}\left(\delta_{qu}-\delta_{qs}\right)$ &  & $e\,p\to e\,K^{0}\Sigma^{+}$  & $-e_{s}\left(\delta_{qd}-\delta_{qs}\right)$ & $-e_{d}\left(\delta_{qd}-\delta_{qs}\right)$\tabularnewline
\cline{1-3} \cline{5-7} 
\end{tabular}

\caption{\label{tab:flavourFactors}Values of the flavor factors $\eta_{q}^{(\pm)}$
for $q=u,\,d,\,s$ quarks. $e_{q}=\left\{ \frac{2}{3},\,-\frac{1}{3},\,-\frac{1}{3}\right\} $
are the electric charges of the quarks. For the sake of reference
in the right column we also placed the corresponding processes on
a proton target which were previously studied in the literature.}
\end{table}

For example, the charged pion ($\pi^{+}$) production on a \emph{proton}
is sensitive only to the combination of GPDs
\begin{equation}
\mathcal{H}_{p\to\pi^{+}n}\left(x,\,\xi,\,t,\,\mu_{F}\right)\equiv\frac{2}{3}\mathcal{H}^{(3)}\left(x,\,\xi,\,t,\,\mu_{F}\right)+\frac{1}{3}\mathcal{H}^{(3)}\left(-x,\,\xi,\,t,\,\mu_{F}\right),
\end{equation}
where $\mathcal{H}^{(3)}\left(x,\,\xi,\,t,\,\mu_{F}\right)\equiv\mathcal{H}^{u}\left(x,\,\xi,\,t,\,\mu_{F}\right)-\mathcal{H}^{d}\left(x,\,\xi,\,t,\,\mu_{F}\right)$,
whereas the charged pion ($\pi^{-}$) production on the \emph{neutron}
has larger sensitivity to the negative-$x$ domain,
\begin{equation}
\mathcal{H}_{n\to\pi^{-}p}\left(x,\,\xi,\,t,\,\mu_{F}\right)\equiv-\frac{2}{3}\mathcal{H}^{(3)}\left(-x,\,\xi,\,t,\,\mu_{F}\right)-\frac{1}{3}\mathcal{H}^{(3)}\left(x,\,\xi,\,t,\,\mu_{F}\right).
\end{equation}
Since $\pi^{-}$ cannot be produced on protons, we believe that the
observation of this process on liquid deuterium target will provide
a clean test of GPD $\mathcal{H}^{(3)}$ at negative $x$. Similarly,
measurement of $\pi^{0}$ on deuterium gives access to a combination
\begin{equation}
\mathcal{H}^{u+d}\left(x,\,\xi,\,t,\,\mu_{F}\right)-\mathcal{H}^{u+d}\left(-x,\,\xi,\,t,\,\mu_{F}\right).
\end{equation}
For processes with strangeness production, the contribution of the
sea quarks is small in JLab kinematics and cancels under the assumption
that the sea is flavor symmetric, for this reason they provide a relatively
clean probes of the GPDs of the valence $u$- and $d$-quarks. We
expect that the process $e\,n\to e\,K^{+}\Sigma^{-}$ should be relatively
easy to access experimentally, since all produced hadrons are charged. 

\subsection{Twist-three corrections}

\label{subsec:Tw3}As was discussed earlier, in modern high-luminosity
experiments a large part of the data come from the region of $Q$
only two or three times larger than nucleon mass $m_{N}$. For this
reason, it is known that in certain channels a significant contribution
comes from the higher-twist contributions due to transversely polarized
photons. This evaluation is very complicated since in the same order
we also have contributions of three-particle correlators which are
completely unknown at present. Under the assumption that these three-particle
distributions are negligible, the result takes the form 
\begin{align}
\delta\mathcal{H}_{\nu'\lambda',\nu\lambda}^{q} & =\left(m_{\nu'\nu}^{q}\delta_{\lambda,-}\delta_{\lambda',+}+n_{\nu'\nu}^{q}\delta_{\lambda,+}\delta_{\lambda',-}\right),
\end{align}
where the coefficients $m_{\pm,\pm}^{q}$ and $n_{\pm,\pm}^{q}$ are
linear combinations of the transversity GPDs, 
\begin{align}
m_{--}^{q} & =\frac{\sqrt{-t'}}{4m}\left[2\tilde{H}_{T}^{q}\,+(1+\xi)E_{T}^{q}-(1+\xi)\tilde{E}_{T}^{q}\right],\\
m_{-+}^{q} & =\sqrt{1-\xi^{2}}\frac{t'}{4m^{2}}\tilde{H}_{T}^{q},\\
m_{+-}^{q} & =\sqrt{1-\xi^{2}}\left[H_{T}^{q}-\frac{\xi^{2}}{1-\xi^{2}}E_{T}^{q}+\frac{\xi}{1-\xi^{2}}\tilde{E}_{T}^{q}-\frac{t'}{4m^{2}}\tilde{H}_{T}^{q}\right],\\
m_{++}^{q} & =\frac{\sqrt{-t'}}{4m}\left[2\tilde{H}_{T}^{q}+(1-\xi)E_{T}^{q}+(1-\xi)\tilde{E}_{T}^{q}\right],
\end{align}

\begin{align}
n_{--}^{q} & =-\frac{\sqrt{-t'}}{4m}\left(2\tilde{H}_{T}^{q}+(1-\xi)E_{T}^{q}+(1-\xi)\tilde{E}_{T}^{q}\right),\\
n_{-+}^{q} & =\sqrt{1-\xi^{2}}\left(H_{T}^{q}-\frac{\xi^{2}}{1-\xi^{2}}E_{T}^{q}+\frac{\xi}{1-\xi^{2}}\tilde{E}_{T}^{q}-\frac{t'}{4m^{2}}\tilde{H}_{T}^{q}\right),\\
n_{+-}^{q} & =\sqrt{1-\xi^{2}}\frac{t'}{4m^{2}}\tilde{H}_{T}^{q},\\
n_{++}^{q} & =-\frac{\sqrt{-t'}}{4m}\left(2\tilde{H}_{T}^{q}+(1+\xi)E_{T}^{q}-(1+\xi)\tilde{E}_{T}^{q}\right),
\end{align}
and we introduced the shorthand notation $t'=-\Delta_{\perp}^{2}/\left(1-\xi^{2}\right)$;
$\Delta_{\perp}=p_{2,\perp}-p_{1,\perp}$ is the transverse part of
the momentum transfer. The coefficient function~(\ref{eq:Coef_function-1})
also gets an additional contribution non-diagonal in parton helicity,

\begin{align}
\delta\mathcal{C}_{\lambda'0,\,\lambda\mu}^{q} & =\left(\delta_{\mu,-}\delta_{\lambda,+}\delta_{\lambda',-}-\delta_{\mu,+}\delta_{\lambda,-}\delta_{\lambda',+}\right)S_{V}^{q}+\mathcal{O}\left(\frac{m^{2}}{Q^{2}}\right),\label{eq:Coef_function-1}
\end{align}
where we introduced shorthand notations
\begin{eqnarray}
S_{V}^{q} & = & \int dz\,\left(\left(\eta_{V+}^{q}c_{+}^{(3,p)}\left(x,\xi\right)+\eta_{V-}^{q}c_{-}^{(3,p)}\left(x,\xi\right)\right)+2\left(\eta_{V+}^{q}c_{+}^{(3,\sigma)}\left(x,\xi\right)-\eta_{V-}^{q}c_{-}^{(3,\sigma)}\left(x,\xi\right)\right)\right),\label{eq:SV_def}
\end{eqnarray}
\begin{equation}
c_{+}^{(3,i)}\left(x,\xi\right)=\frac{4\pi i\alpha_{s}f_{\pi}\xi}{9\,Q^{2}}\int_{0}^{1}dz\frac{\phi_{3,i}(z)}{z(x+\xi)^{2}},\quad c_{-}^{(3,i)}\left(x,\xi\right)=\frac{4\pi i\alpha_{s}f_{\pi}\xi}{9\,Q^{2}}\int_{0}^{1}dz\frac{\phi_{3,i}(z)}{(1-z)(x-\xi)^{2}};\label{eq:Tw3_coefFunction}
\end{equation}
and the twist-three pion distributions are defined as 
\begin{eqnarray}
\phi_{3}^{(p)}\left(z\right) & = & \frac{1}{f_{\pi}\sqrt{2}}\frac{m_{u}+m_{d}}{m_{\pi}^{2}}\int\frac{du}{2\pi}e^{i(z-0.5)u}\left\langle 0\left|\bar{\psi}\left(-\frac{u}{2}n\right)\gamma_{5}\psi\left(\frac{u}{2}n\right)\right|\pi(q)\right\rangle ,\label{eq:DA3p}
\end{eqnarray}

\begin{eqnarray}
\phi_{3}^{(\sigma)}\left(z\right) & = & \frac{3i}{\sqrt{2}f_{\pi}}\frac{m_{u}+m_{d}}{m_{\pi}^{2}}\int\frac{du}{2\pi}e^{i(z-0.5)u}\left\langle 0\left|\bar{\psi}\left(-\frac{u}{2}n\right)\sigma_{+-}\gamma_{5}\psi\left(\frac{u}{2}n\right)\right|\pi(q)\right\rangle .\label{eq:DA3s}
\end{eqnarray}

Thanks to the symmetry of $\phi_{p}$ and the antisymmetry of $\phi_{\sigma}$
with respect to charge conjugation, the dependence on the pion DAs
factorizes in the collinear approximation and contributes only as
the minus first moment of the linear combination of the twist-3 DAs,
$\phi_{p}(z)+2\phi_{\sigma}(z)$, 
\begin{equation}
\left\langle \phi_{3}^{-1}\right\rangle =\int_{0}^{1}dz\frac{\phi_{3}^{(p)}\left(z\right)+2\phi_{3}^{(\sigma)}\left(z\right)}{z}.
\end{equation}
In the general case the coefficient function ~(\ref{eq:Tw3_coefFunction})
leads to collinear divergencies near the points $x=\pm\xi$ when substituted
to (\ref{eq:MConv}). As was noted in~\cite{Goloskokov:2009ia},
this singularity is naturally regularized by maintaining the small
transverse momentum of the quarks inside the meson. Such regularization
modifies~(\ref{eq:Tw3_coefFunction}) to 
\begin{align}
c_{+}^{(3,i)}\left(x,\xi\right) & =\frac{4\pi i\alpha_{s}f_{\pi}\xi}{9\,Q^{2}}\int_{0}^{1}dz\,d^{2}l_{\perp}\frac{\phi_{3,i}\left(z,\,\ell_{\perp}\right)}{(x+\xi-i0)\left(z\,(x+\xi)+\frac{2\xi\,\ell_{\perp}^{2}}{Q^{2}}\right)},\label{eq:c3Plus}\\
c_{-}^{(3,i)}\left(x,\xi\right) & =\frac{4\pi i\alpha_{s}f_{\pi}\xi}{9\,Q^{2}}\int_{0}^{1}dz\,d^{2}l_{\perp}\frac{\phi_{3,i}\left(z,\,\ell_{\perp}\right)}{(x-\xi+i0)\left((1-z)(x-\xi)-\frac{2\xi\,\ell_{\perp}^{2}}{Q^{2}}\right)},\label{eq:c3Minus}
\end{align}
where $\ell_{\perp}$ is the transverse momentum of the quark, and
we tacitly assume absence of any other transverse momenta in the coefficient
function. 

\section{Results and discussion}

\label{sec:Results}

In this section we would like to present numerical results for charged
current pion production. For the sake of definiteness, for numerical
estimates we use the Kroll-Goloskokov parametrization of GPDs~\cite{Goloskokov:2006hr,Goloskokov:2007nt,Goloskokov:2008ib,Goloskokov:2009ia,Goloskokov:2011rd}.
We would like to comment briefly on the inclusion of the so-called
$t$-channel pion pole. As was shown long ago, this contribution can
be incorporated into the GPD $\tilde{E}$,~\cite{Vanderhaeghen:1999xj,Penttinen:1999th,Goloskokov:2009ia}
and gives the dominant contribution at small-$t$,
\begin{align}
\tilde{E}_{({\rm pole})}^{u}\left(x,\,\xi,\,t\right)=-\tilde{E}_{({\rm pole})}^{d}\left(x,\,\xi,\,t\right) & =\theta\left(|x|<\xi\right)\Phi_{\pi}\left(\frac{x+\xi}{2\xi}\right)\frac{F_{P}\left(t\right)}{4\xi},\label{eq:PionPole}\\
F_{P}(t) & =\frac{2\sqrt{2}m\,f_{\pi}g_{\pi NN}}{t-m_{\pi}^{2}}\,\frac{\Lambda_{N}^{2}-m_{\pi}^{2}}{\Lambda_{N}^{2}-t'}
\end{align}
where $t'\equiv t-t_{{\rm min}}=-\Delta_{\perp}^{2}/\left(1-\xi^{2}\right),$
and following~\cite{Goloskokov:2009ia} we use constants $g_{\pi NN}\approx13.6,\,\Lambda_{N}\approx0.51\,{\rm GeV}$.
However, when we consider kaon production, we expect that the pole
should be located at $t\sim m_{K}^{2}$ rather than at $t\sim m_{\pi}^{2}$.
Physically, this contradiction signals that the $SU(3)$-based symmetry
relations, which relate the transition GPDs $N\to Y$ to GPDs of the
proton might require modification near $|t|\lesssim m_{K}^{2}$. In
what follows, we make proper adjustments of the pole term in transition
GPDs, e.g. for $K^{+}$ production we use~\cite{Goeke:2001tz,Penttinen:1999th}
\begin{align}
\tilde{E}_{({\rm pole})}^{u}\left(x,\,\xi,\,t\right)=-\tilde{E}_{({\rm pole})}^{s}\left(x,\,\xi,\,t\right) & =\theta\left(|x|<\xi\right)\Phi_{K}\left(\frac{x+\xi}{2\xi}\right)\frac{F_{P,K}\left(t\right)}{4\xi},\label{eq:KaonPole}\\
F_{P,K}(t) & =\frac{2\sqrt{2}m\,f_{K}g_{KNY}}{t-m_{K}^{2}}\frac{\Lambda_{N}^{2}-m_{K}^{2}}{\Lambda_{N}^{2}-t'}.
\end{align}

For numerical estimates, we assume that $g_{KNY}\approx g_{\pi NN}$.
We expect that the kaon pole contribution should be more suppressed
than the pion pole contribution. For both the pion and kaon leading-twist
wave function we use the asymptotic form,~$\phi_{2}(z)=6\,z\,\left(1-z\right)$,
and due to the symmetry $z\to1-z$ the contributions~(\ref{eq:PionPole},\ref{eq:KaonPole})
are even functions of the variable $x$. For estimates of the twist-3
contribution introduced in Section~\ref{sec:DVMP_Xsec}, we use the
parametrization suggested in~\cite{Goloskokov:2009ia,Goloskokov:2011rd},
\begin{align}
\phi_{3}\left(z,\,l_{\perp}\right) & =\phi_{3;p}\left(z,\,l_{\perp}\right)+2\phi_{3;\sigma}\left(z,\,l_{\perp}\right)=\frac{16\pi^{3/2}}{\sqrt{6}}f_{\pi}a_{\pi}^{3}l_{\perp}\phi_{as}(z)\exp\left(-a_{p}^{2}l_{\perp}^{2}\right),\label{eq:phi_3}\\
a_{\pi} & =\left[\sqrt{8}\pi f_{\pi}\right]^{-1},\quad a_{p}\approx2\,{\rm GeV}^{-1}.
\end{align}

\begin{figure}
\includegraphics[width=9cm]{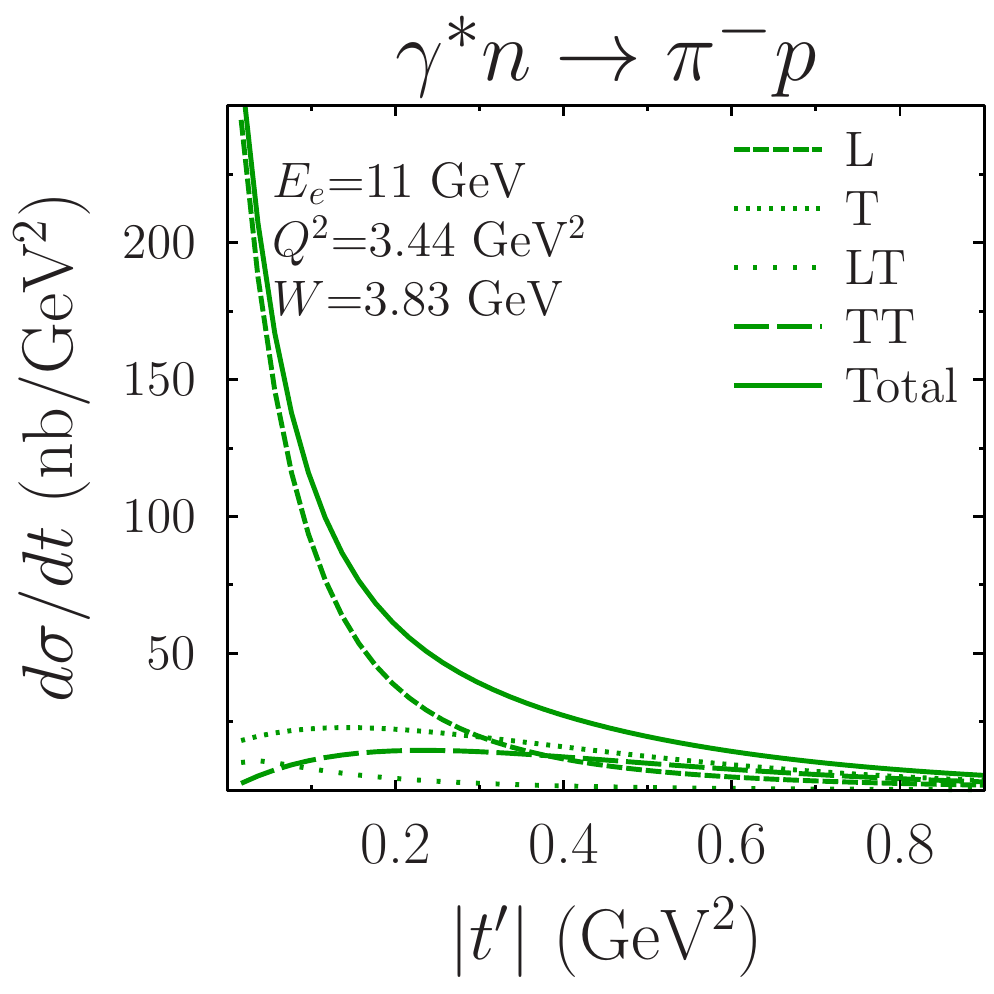}\includegraphics[width=9cm]{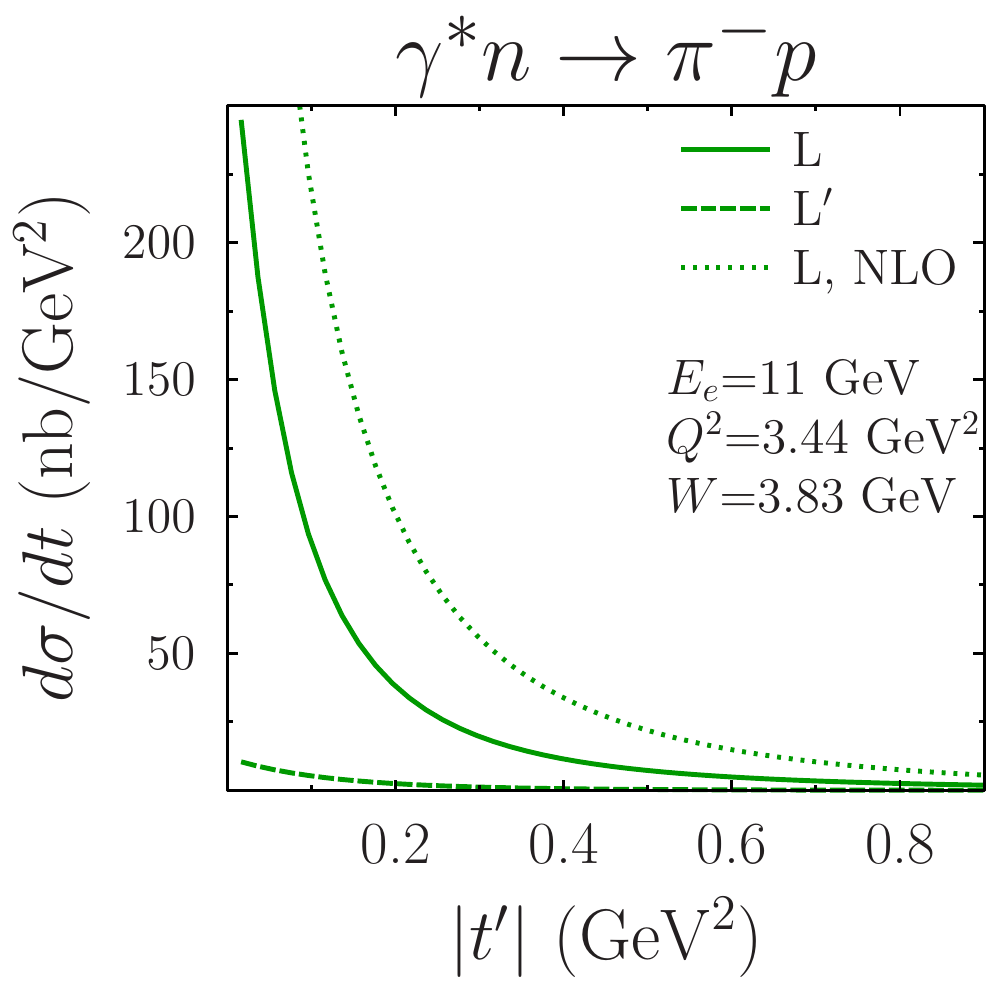}

\protect\caption{\label{fig:DVMP-piMinus}(color online) Left: Different components
of charged pion production on a neutron, $e\,n\to e\,\pi^{-}p$. The
dominant contribution to the total cross-section (solid line) comes
from the longitudinal cross-section $d\sigma_{L}/dt$ due to the $t$-channel
pion pole. At larger $t$ the contributions of transversity GPDs become
comparable to the leading-twist contribution $d\sigma_{L}/dt.$ Right:
comparison of leading twist cross-sections, taking into account the
pole term (solid curve, marked with symbol ``$L$'') and without
it (dashed curve, marked with symbol ``$L'$''). Also, in the same
plot we have shown the NLO corrections to the coefficient functions.
For the ease of comparison, in both plots we've chosen the same values
of $W,\,Q^{2}$ as in~\cite{Goloskokov:2009ia}. The value of Bjorken
$x_{B}$ is $x_{B}\approx0.2$.}
\end{figure}

We would like to start our discussion with the process $en\to e\pi^{-}p$,
which is sensitive to the GPD $\mathcal{H}^{(3)}=\mathcal{H}^{(u)}-\mathcal{H}^{(d)}$.
As was indicated in~Section~\ref{subsec:LeadingTwist}, this process
probes the same GPD as in the case of the charged pion production
on the proton ($ep\to e\pi^{+}n$), but is expected to have stronger
sensitivity to the region of negative $x$. In Figure~\ref{fig:DVMP-piMinus}
we have shown different components of the cross-section defined in
Eq.~(\ref{eq:sigma_def-1-1}). Just as in the proton case, the cross-section
of this process at small $t$ is dominated by the $t$-channel pion
pole, and since the corresponding pole contribution to the GPD $\tilde{E}$
is symmetric w.r.t. $x\leftrightarrow-x$, as could be seen from the
Table~\ref{tab:flavourFactors}, its contribution to the amplitudes
of $ep\to e\pi^{+}n$ and $en\to e\pi^{-}p$ processes differs only
by sign. At larger values of $t$, the pion pole contribution fades
out, the contribution of the transverse cross-section becomes more
pronounced and eventually becomes comparable to $d\sigma_{L}/dt$.
In the right panel of the Figure~\ref{fig:DVMP-piMinus} we compare
the values of the cross-section $d\sigma_{L}/dt$ with and without
pole contributions (solid and dashed lines respectively) and demonstrate
the importance of the pion pole contribution. Also we've shown with
dashed line the result when the NLO corrections to the coefficient
functions are taken into account. As we can see, the corrections are
large, and similar conclusions are valid for other processes mentioned
below. For this reason we believe that the GPD analysis from data
done in~\cite{Goloskokov:2009ia,Goloskokov:2011rd} should be repeated
taking into account NLO corrections. We expect that the process $en\to e\pi^{-}p$
could be relatively easy to access experimentally on deuterium target,
since all final hadrons are charged, and $\pi^{-}$ can be produced
only on protons. It is also very interesting to measure the ratio
of the cross-sections $d\sigma_{L}/dt$ in $ep\to e\pi^{+}n$ and
$en\to e\pi^{-}p$ processes: in the case of pion pole dominance it
is expected that this ratio should be close to unity.

\begin{figure}
\includegraphics[width=9cm]{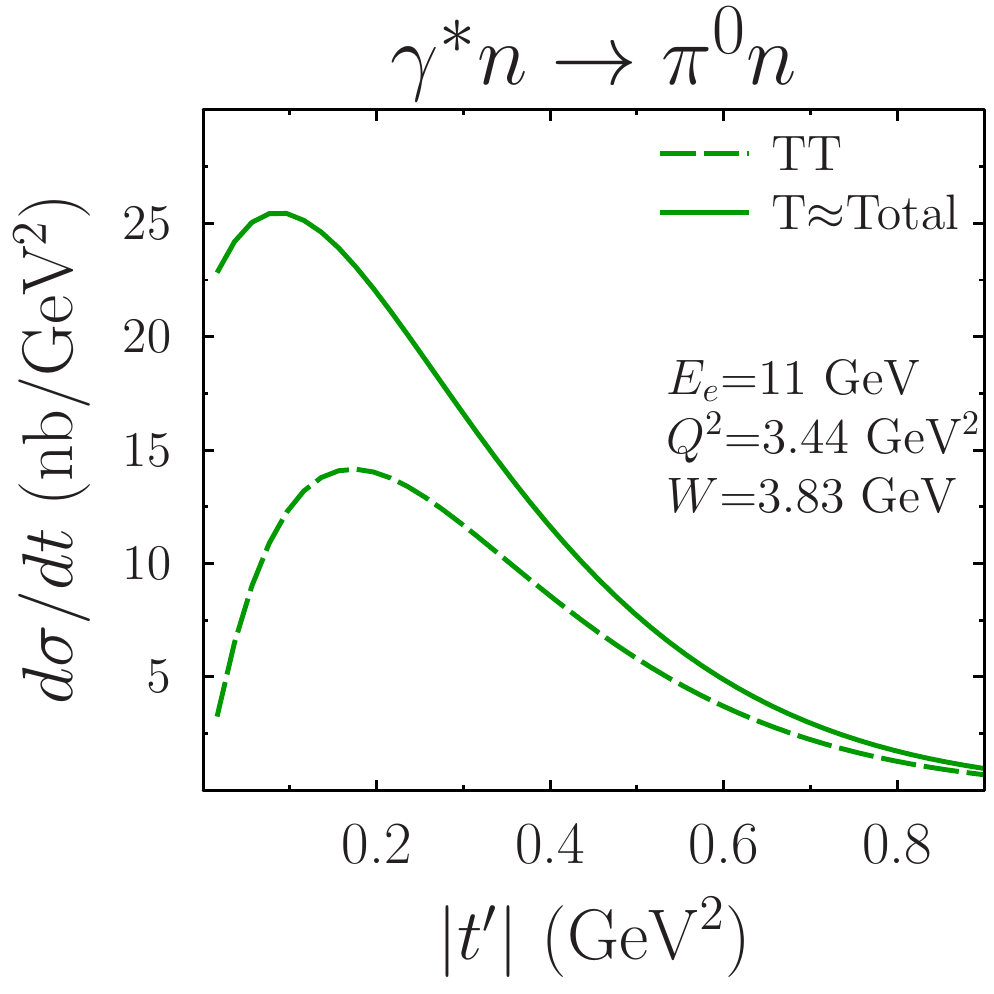}\includegraphics[width=9cm]{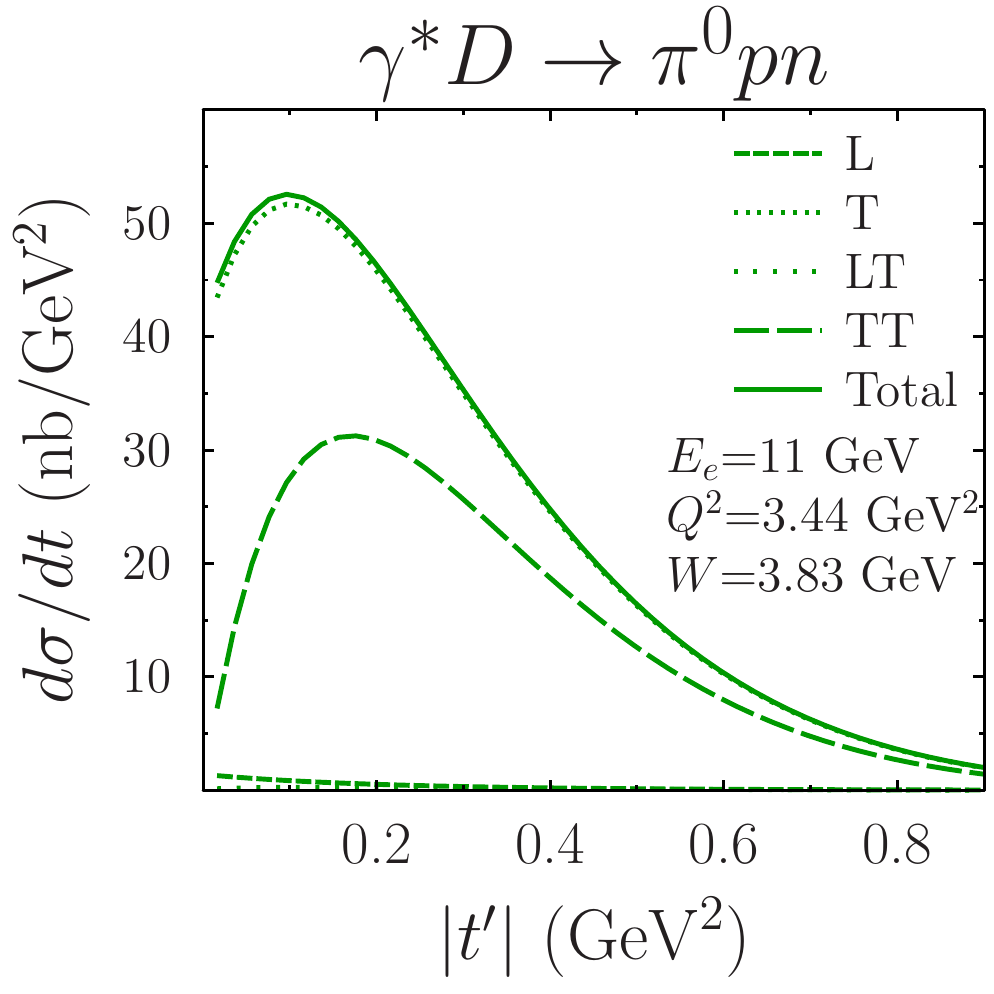}

\protect\caption{\label{fig:DVMP-piZero}(color online) Left: Different components
of neutral pion production on the neutron, $e\,n\to e\,\pi^{0}p$.
The dominant contribution to the total cross-section (solid line)
comes from the transverse cross-section $d\sigma_{T}/dt$. Other components
(not shown) are negligible. Right: The same plot for the incoherent
process on the deuteron, $eD\to e\pi^{0}pn$. For ease of comparison,
in both plots we've chosen the same values of $W,\,Q^{2}$ as in~\cite{Goloskokov:2009ia}.
The value of the Bjorken $x_{B}$ is $x_{B}\approx0.2$.}
\end{figure}

In the Figure~\ref{fig:DVMP-piZero} we show the cross-sections for
the case of neutral pions. The contribution of the pion pole~(\ref{eq:PionPole})
in the amplitude exactly cancels in this case due to the symmetry
of~(\ref{eq:PionPole}) w.r.t. $x\leftrightarrow-x$ and the structure
of the flavor factors (see Table~\ref{tab:flavourFactors}). For
this reason the cross-section $d\sigma_{L}/dt$ is significantly smaller,
and the cross-section is dominated by the transverse terms. The neutral
pion production ($en\to e\pi^{0}n$) has the same final state as in
proton case and is experimentally indistinguishable from it. Experimentally,
the $\pi^{0}$ production on neutrons might be measured either in
coherent processes $e\,A\to e\,A\,\pi^{0}$ or in incoherent processes
$e\,A\to e\,n\,X.$ The contributions of the coherent process peaks
at small-$t$ and significantly depends on the implemented model of
nuclear structure~\cite{Guzey:2005ba}. In contrast, the contribution
of incoherent process has a milder dependence on nuclear effects and
is not suppressed at moderate $t$. The cleanest process for such
study is the production on a deuterium target, $eD\to e\pi^{0}pn$,
which has negligibly small nuclear effects due to very weak binding
of deuteron. The feasibility of such measurements has been demonstrated
recently at Hall A~\cite{Mazouz:2017skh,Bosted:2016leu}, although
due to the low energy of the incident electron beam the virtuality
$Q^{2}$ was too low for its consideration in the Bjorken limit. The
cross-section for the process $eD\to e\pi^{0}pn$ is shown for the
sake of reference in the right panel of the Figure~\ref{fig:DVMP-piZero}.
As could be seen from the plot, it gets the dominant contribution
from the transversity GPDs.

\begin{figure}
\includegraphics[width=6cm]{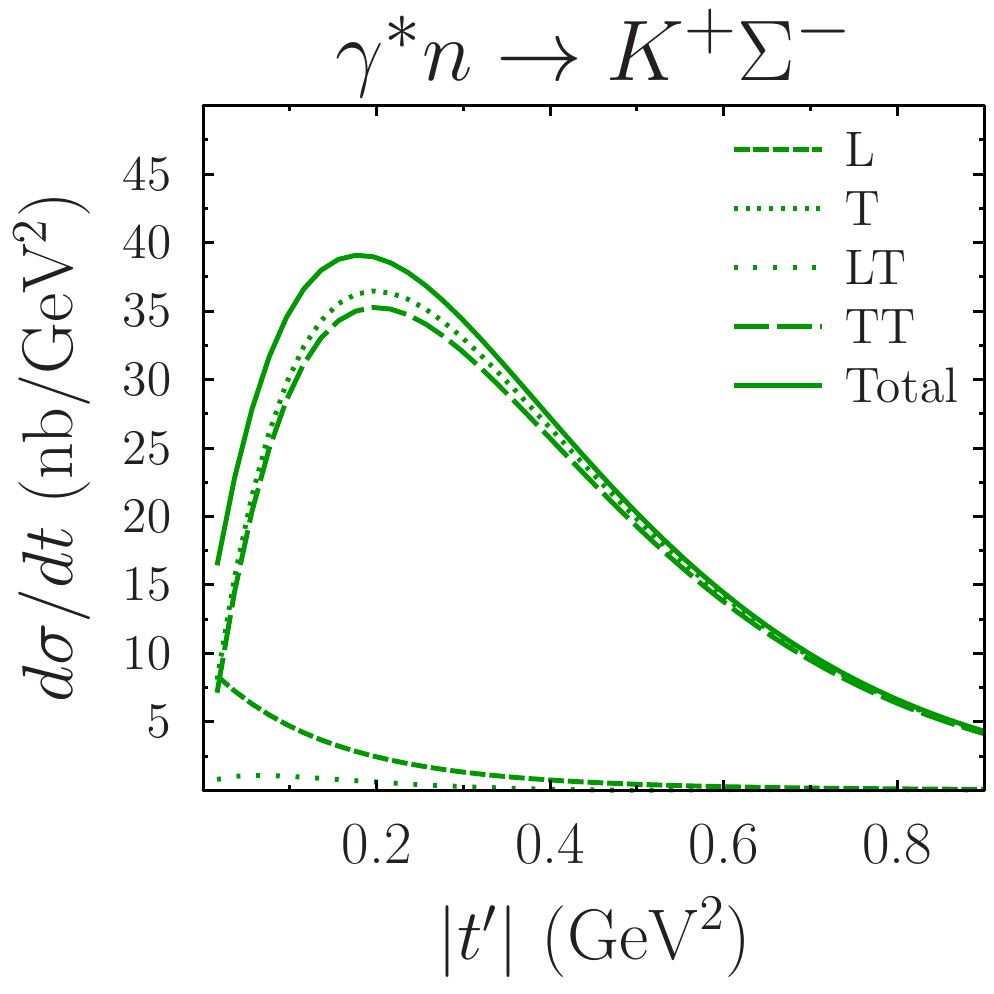}\includegraphics[width=6cm]{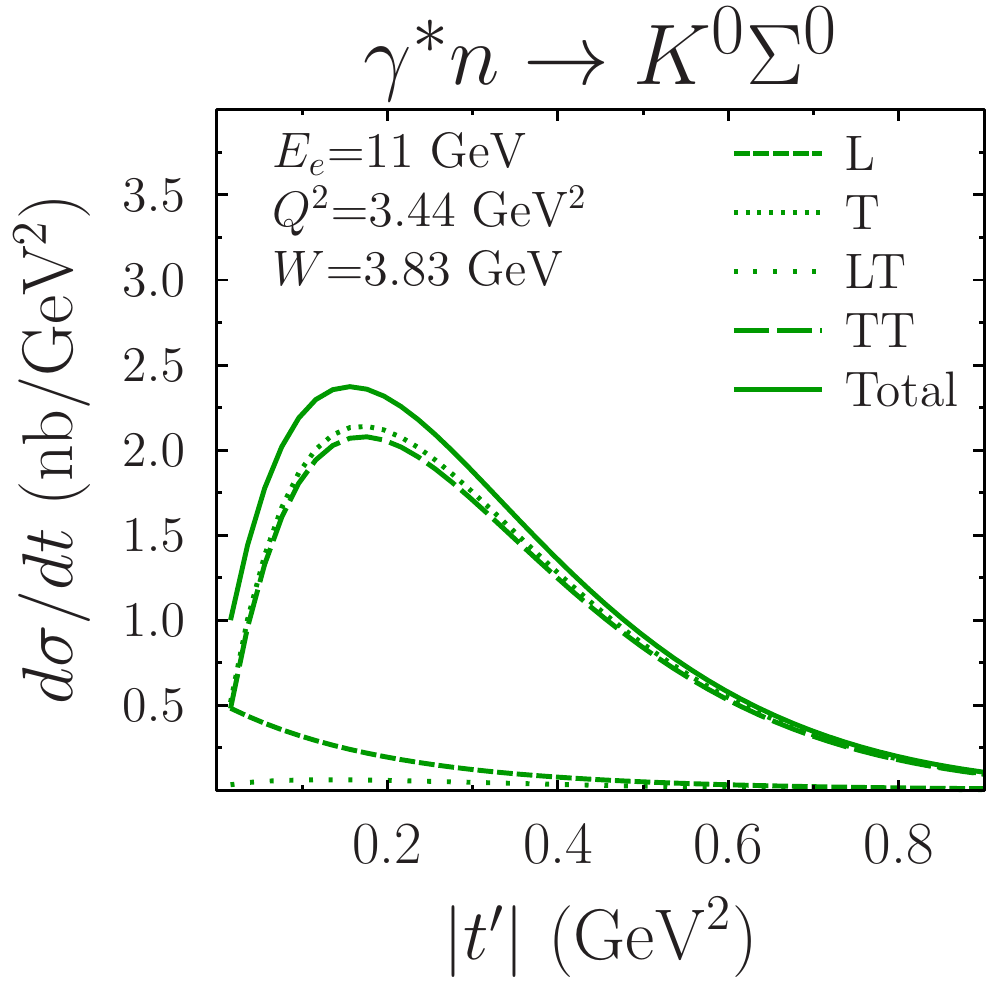}\includegraphics[width=6cm]{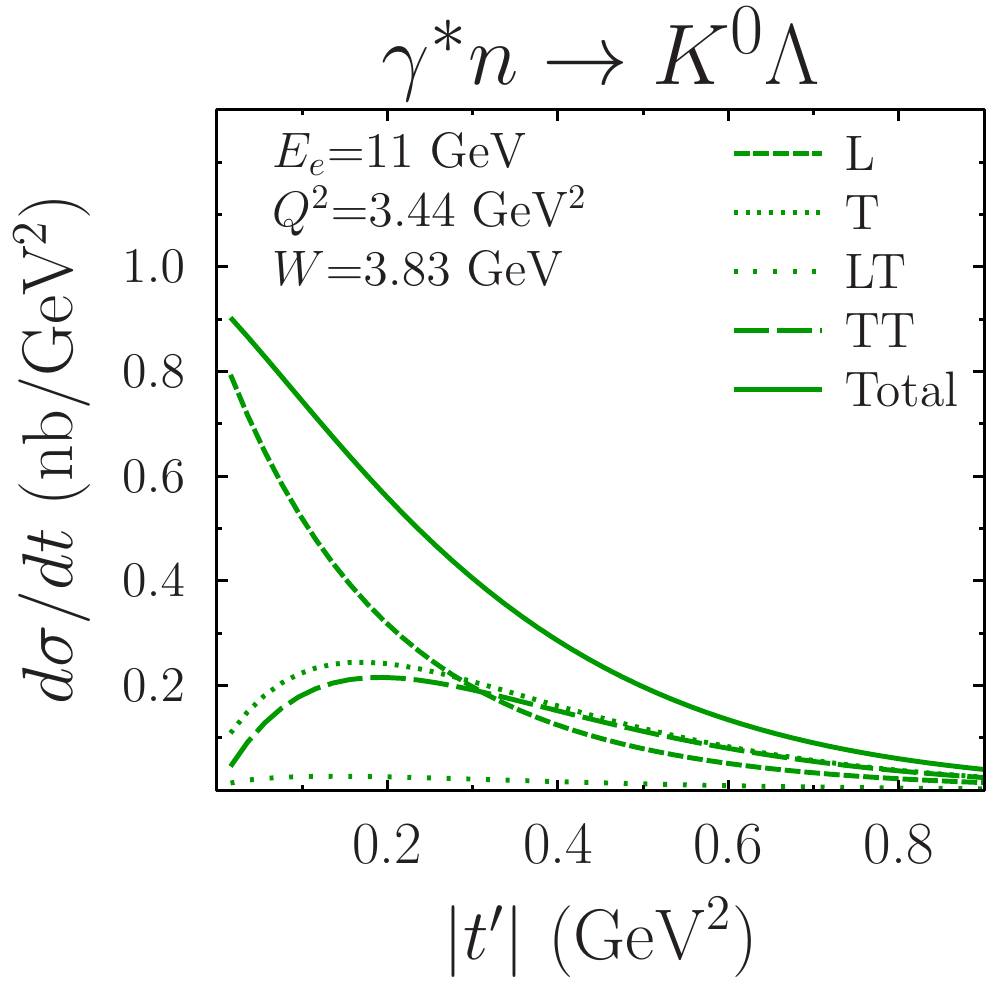}

\protect\caption{\label{fig:DVMP-kaons}(color online) Cross-sections for the strangeness
production on the neutron. In case of $\gamma^{*}n\to K^{+}\Sigma^{-}$
and $\gamma^{*}n\to K^{0}\Sigma^{0}$ the dominant contribution to
the total cross-section (solid line) comes from the transverse cross-section
$d\sigma_{T}/dt$. In the process $\gamma^{*}n\to K^{0}\Lambda$ the
dominant contribution comes from $d\sigma_{L}/dt$ (dashed line) which
is mildly enhanced at small-$t'$ due to kaon pole~(\ref{eq:KaonPole}).
For the ease of comparison, in both plots we've chosen the same values
of $W,\,Q^{2}$ as in~\cite{Goloskokov:2009ia}. The value of the
Bjorken $x_{B}$ is $x_{B}\approx0.2$.}
\end{figure}

Finally, in Figure~\ref{fig:DVMP-kaons} we show predictions for
the differential cross-section of kaon production ($K^{+}$and $K^{0}$).
The sea quark densities are small in the kinematics of JLab, and additionally,
as could be seen from the structure of the flavor factors in Table~\ref{tab:flavourFactors},
the contribution of the sea quarks cancels in the case of the flavor
symmetric sea, as implemented in the parametrization~\cite{Goloskokov:2009ia,Goloskokov:2011rd},
and thus kaon production is sensitive to the valence $u$- and $d$-quarks.
The largest cross-section has the process $e\,n\to e\,K^{+}\Sigma^{-}$,
which gets the dominant contribution from the transversity GPDs of
the $u$-quarks. We expect that experimentally it should be very easy
to access it since both produced hadrons are charged, and additionally
there is no interference from protons. The process $ep\to e'K^{0}\Sigma^{0}$is
sensitive to the same valence $u$-quarks, however, as can be seen
from the flavor factors in Table~\ref{tab:flavourFactors}, it is
suppressed compared to $K^{+}\Sigma^{-}$ by a relative factor $\sim(e_{d}/e_{u})^{2}/2$
which results in approximately an order of magnitude smaller cross-sections.
Finally, the process $en\to eK^{0}\Lambda$ has the smallest cross-section
due to the suppression by a relative factor $\sim(e_{d}/e_{u})^{2}/6$
and additionally is suppressed due to very specific combination of
GPDs $2\mathcal{H}^{d}-\mathcal{H}^{u}$ which contribute to it, so
we believe that it will be challenging to access it experimentally. 

\section{Conclusions}

In this paper we analyzed the production of pions and kaons on neutron
targets using the modern GPD parametrization~\cite{Goloskokov:2006hr,Goloskokov:2007nt,Goloskokov:2008ib,Goloskokov:2009ia,Goloskokov:2011rd}.
We estimated the cross-sections in the kinematics of upgraded 12 GeV
Jefferson Laboratory experiments, assuming that the measurements will
be done on liquid deuterium targets. We found that four processes
($\gamma^{*}n\to\pi^{-}p,\,\,K^{+}\Sigma^{-},\,\,K^{0}\Sigma^{+},\,\,K^{0}\Lambda$)
might proceed only on neutrons and thus provide clean probes of the
corresponding GPDs combinations. They do not get contributions from
gluons nor from sea quarks (provided the sea is flavor symmetric),
and thus probe valence GPDs of $u$- and $d$-quarks. The neutral
pion production ($\gamma^{*}N\to\pi^{0}N$) gets comparable contributions
from both proton and neutrons and thus is more challenging for the
extraction of GPDs. The cross-sections of pion and charged kaon production
are sufficiently large and comparable to the corresponding processes
on the protons. Additionally, in the case of charged pion and kaon
production all produced hadrons are charged, which facilitates the
reconstruction of the kinematics of the process and allows measurements
with reasonable statistics. The code for evaluation of the cross-sections
with arbitrary GPD models is available from the authors on demand.

\section*{Acknowledgments}

This research was partially supported by Proyecto Basal FB 0821 (Chile),
the Fondecyt (Chile) grants 1180232 and 1140377, CONICYT (Chile) grant
PIA ACT1413. Powered@NLHPC: This research was partially supported
by the supercomputing infrastructure of the NLHPC (ECM-02). Also,
we thank Yuri Ivanov for technical support of the USM HPC cluster
where part of evaluations were done.

\appendix

\section{NLO coefficient function}

\label{sec:NLOCoef}The function $T^{(1)}\left(v,\,z\right)$ in~(\ref{eq:c2-1})
encodes NLO corrections to the coefficient function. 

Explicitly, this function is given by 
\begin{align}
T^{(1)}\left(v,\,z\right) & =\frac{1}{2vz}\left[\frac{4}{3}\left([3+\ln(v\,z)]\,\ln\left(\frac{Q^{2}}{\mu_{F}^{2}}\right)+\frac{1}{2}\ln^{2}\left(v\,z\right)+3\ln(v\,z)-\frac{\ln\bar{v}}{2\bar{v}}-\frac{\ln\bar{z}}{2\bar{z}}-\frac{14}{3}\right)\right.\label{eq:T1-1}\\
 & +\beta_{0}\left(\frac{5}{3}-\ln(v\,z)-\ln\left(\frac{Q^{2}}{\mu_{R}^{2}}\right)\right)\nonumber \\
 & -\frac{1}{6}\left(2\frac{\bar{v}\,v^{2}+\bar{z}\,z^{2}}{(v-z)^{3}}\left[{\rm Li}_{2}(\bar{z})-{\rm Li}_{2}(\bar{v})+{\rm Li}_{2}(v)-{\rm Li}_{2}(z)+\ln\bar{v}\,\ln z-\ln\bar{z}\,\ln v\right]\right.\nonumber \\
 & +2\frac{v+z-2v\,z}{(v-z)^{2}}\ln\left(\bar{v}\bar{z}\right)+2\left[{\rm Li}_{2}(\bar{z})+{\rm Li}_{2}(\bar{v})-{\rm Li}_{2}(z)-{\rm Li}_{2}(v)+\ln\bar{v}\,\ln z+\ln\bar{z}\,\ln v\right]\nonumber \\
 & +\left.\left.4\frac{v\,z\,\ln(v\,z)}{(v-z)^{2}}-4\ln\bar{v}\,\ln\bar{z}-\frac{20}{3}\right)\right],\nonumber 
\end{align}
where $\beta_{0}=\frac{11}{3}N_{c}-\frac{2}{3}N_{f}$, ${\rm Li}_{2}(z)$
is the dilogarithm function, and $\mu_{R}$ and $\mu_{F}$ are the
renormalization and factorization scales respectively. For the vector
meson production in processes when the internal state of the hadron
is not changed, the additional contribution comes from gluons and
singlet (sea) quarks~\cite{Belitsky:2001nq,Ivanov:2004zv,Diehl:2007hd}~\footnote{For the sake of simplicity, we follow~\cite{Diehl:2007hd} and assume
that the factorization scale $\mu_{F}$ is the same for both the generalized
parton distribution and the pion distribution amplitude.}, 

Some coefficient functions~ have non-analytic behavior $\sim\ln^{2}v$
for small $v\approx0$ ($x=\pm\xi\mp i0$), which signals that the
collinear approximation might be not valid near this point. This singularity
in the collinear limit occurs due to the omission of the small transverse
momentum $l_{M,\perp}$ of the quark inside a meson~\cite{Goloskokov:2009ia}.
For this reason the contribution of the region $|v|\sim l_{M,\perp}^{2}/Q^{2}$
for finite $Q^{2}$ (below the Bjorken limit) should be treated with
due care. However, a full evaluation of $T^{(1)}\left(v,\,z\right)$
beyond the collinear approximation (taking into account all higher
twist corrections) presents a challenging problem and has not been
done so far. It was observed in~\cite{Diehl:2007hd}, that the singular
terms might be eliminated by a redefinition of the renormalization
scale $\mu_{R}$, however near the point $v\approx0$ the scale $\mu_{R}^{2}$
becomes soft, $\mu_{R}^{2}\sim z\,v\,Q^{2}\lesssim l_{\perp}^{2}$
which is another manifestation that nonperturbative effects become
relevant. For this reason, sufficiently large value of $Q^{2}$ should
be used to mitigate contributions of higher twist effects. As we will
see below, for $Q^{2}\approx4$ GeV$^{2}$ the contribution of this
soft region is small, so the collinear factorization is reliable.

 \end{document}